\documentstyle[aps,prl,epsfig,float,twocolumn]{revtex}
\begin{document}
\twocolumn[\hsize\textwidth\columnwidth\hsize\csname
@twocolumnfalse\endcsname

\title{Zitterbewegung of electronic wave packets in semiconductor 
nanostructures}

\author{John Schliemann$^{1}$, Daniel Loss$^{1}$, and R.~M. Westervelt$^{2}$}

\address{$^{1}$Department of Physics and Astronomy, University of Basel,
CH-4056 Basel, Switzerland\\
$^{2}$Division of Engineering and Applied Sciences, Harvard University, 
Cambridge, Massachusetts 02138, USA}

\date{\today}

\maketitle

\begin{abstract}
We study the zitterbewegung of electronic wave packets in 
III-V zinc-blende semiconductor quantum wells due to spin-orbit coupling. 
Our results suggest a direct experimental proof of this fundamental effect,
confirming a long-standing theoretical prediction. For electron motion in
a harmonic quantum wire, we numerically and analytically 
find a resonance condition maximizing the zitterbewegung. 
\end{abstract}
\vskip2pc]

The emerging field of spintronics has generated a large deal of interest in
the effects  of spin-orbit coupling in semiconductor
nanostructures and their possible applications \cite{overview}. Spin-orbit
coupling is a relativistic effect described by the Dirac equation and its
nonrelativistic expansion \cite{Feshbach58}. In semiconductors spin-orbit 
coupling of itinerant electrons is much stronger than in vacuum. This is
due to the interplay of crystal symmetry and the strong electric fields 
of the atom cores \cite{Rashba04}. In fact, the effective models describing
the band structure of III-V semiconductors have many similarities to the
Dirac equation. However, the fundamental gap between conduction and valence 
band in a semicondcutor is of
order an electron volt or less, which is much smaller than the gap between
solutions of positive and negative energy of the free Dirac equation.
This observation can serve as a general heuristic explanation for the
importance of spin-orbit effects in semiconductors \cite{Rashba04}. 

Another prediction of relativistic quantum mechanics is the zitterbewegung 
of electrons \cite{Feshbach58}, which, however, has not been observed 
experimentally yet. For free electrons, i.e. in the
absence of an external potential, 
such an oscillatory motion occurs if 
solutions of both positive and negative energy 
of the free Dirac equation have a finite 
weight in a given quantum state. In this letter we investigate the 
zitterbewegung of electron wave packets under the influence of the enhanced
spin-orbit coupling in III-V zinc-blende semiconductor quantum wells.
Strong spin-orbit coupling generally requires large gradients of the
external potential, as they are provided by the 
heavy atom cores in such systems.

An important effective
contribution to spin-orbit coupling in such systems is the
Rashba term which is due to structure-inversion asymmetry of the confining
potential and takes the following form \cite{Rashba60}
\begin{equation}
{\cal
H}_{R}=(\alpha/\hbar)\left(p_{x}\sigma^{y}-p_{y}\sigma^{x}\right),
\label{rashba}
\end{equation}
where $\vec p$ is the momentum of the electron confined in a
two-dimensional geometry, and $\vec\sigma$ the vector of Pauli
matrices. The Rashba coefficient $\alpha$ is 
essentially proportional to the
potential gradient across the well and therefore tunable by an external gate.
Thus, the single-particle
Hamiltonian is given by ${\cal H}=\vec p^{2}/2m+ {\cal H}_{R}$, where $m$
is the effective band mass.
The components of the time-dependent position operator
\begin{equation}
\vec r_{H}(t)=e^{i{\cal H}/\hbar}\vec r(0)e^{-i{\cal H}/\hbar}
\end{equation}
in the Heisenberg picture read explicitly
\begin{eqnarray}
x_{H}(t) & = & x(0)+ \frac{p_{x}}{m}t
+\frac{p_{y}}{p^{2}}\frac{\hbar}{2}\sigma^{z}
\left(1-\cos\left(\frac{2\alpha p}{\hbar^{2}}t\right)\right)\nonumber\\
 & + & \frac{p_{x}}{p^{3}}\frac{\hbar}{2}
\left(p_{x}\sigma^{y}-p_{y}\sigma^{x}\right)
\left(\frac{2\alpha p}{\hbar^{2}}t
-\sin\left(\frac{2\alpha p}{\hbar^{2}}t\right)\right)\nonumber\\
 & + & \frac{1}{p}\frac{\hbar}{2}\sigma^{y}
\sin\left(\frac{2\alpha p}{\hbar^{2}}t\right)\,,\\
y_{H}(t) & = & y(0)+ \frac{p_{y}}{m}t
-\frac{p_{x}}{p^{2}}\frac{\hbar}{2}\sigma^{z}
\left(1-\cos\left(\frac{2\alpha p}{\hbar^{2}}t\right)\right)\nonumber\\
 & + & \frac{p_{y}}{p^{3}}\frac{\hbar}{2}
\left(p_{x}\sigma^{y}-p_{y}\sigma^{x}\right)
\left(\frac{2\alpha p}{\hbar^{2}}t
-\sin\left(\frac{2\alpha p}{\hbar^{2}}t\right)\right)\nonumber\\
 & - & \frac{1}{p}\frac{\hbar}{2}\sigma^{x}
\sin\left(\frac{2\alpha p}{\hbar^{2}}t\right)\,,
\end{eqnarray}
where the operators $\vec p$ and $\vec\sigma$ on the right hand sides are
in the Schr\"odinger picture and therefore time-independent.

We now proceed by evaluating the above time-dependent position operators 
within a Gaussian wave packet with initial spin polarization along the 
$z$-direction perpendicular to the quantum well,
\begin{equation}
\langle\vec r|\psi\rangle=\frac{1}{2\pi}\frac{d}{\sqrt{\pi}}
\int d^{2}k\,e^{-\frac{1}{2}d^{2}\left(\vec k-\vec k_{0}\right)^{2}}
e^{i\vec k\vec r}
\left(
\begin{array}{c}
1 \\ 0
\end{array}
\right)\,.
\end{equation}
Clearly we have $\langle\psi|\vec r|\psi\rangle=0$,
$\langle\psi|\vec p|\psi\rangle=\hbar\vec k_{0}$, and the variances 
of the position and momentum operators are 
$\left(\Delta x\right)^{2}=\left(\Delta y\right)^{2}=d^{2}/2$,
$\left(\Delta p_{x}\right)^{2}=\left(\Delta p_{y}\right)^{2}
=\hbar^{2}/2d^{2}$.
Thus, the group velocity of the wave packet is given by 
$\hbar\vec k_{0}/m$ , while 
its spatial width is described by the parameter $d$ with the minimum
uncertainty product typical for Gaussian wave packets,
$\Delta p_{x}\Delta x=\Delta p_{y}\Delta y=\hbar/2$.

A direct calculation gives
\begin{eqnarray}
\langle\psi|x_{H}(t)|\psi\rangle & = & \frac{\hbar k_{0x}}{m}
+\frac{d}{2\pi}e^{-d^{2}k_{0}^{2}}\int_{0}^{2\pi}d\varphi\sin\varphi
\nonumber\\
& & \cdot\int_{0}^{\infty}dq\,e^{-q^{2}+2dq
\left(k_{0x}\cos\varphi+k_{0y}\sin\varphi\right)}\nonumber\\
& & \qquad\cdot\left(1-\cos\left(\frac{2\alpha q}{\hbar d}t\right)\right)\,.
\end{eqnarray}
In the above expression, $\varphi$ is a usual polar angle in the $xy$-plane,
and $q$ is a dimensionless integration variable.
The time dependence in the integral can be viewed as a zitterbewegung the 
electron performs under the influence of spin-orbit coupling. Clearly, this
integral contribution vanishes for $k_{0y}=0$, i.e. if the group velocity
is along the $x$-direction. More generally, one finds that
\begin{equation}
\langle\psi|\vec k_{0}\cdot\vec r_{H}(t)/k_{0}|\psi\rangle
=\hbar k_{0}t/m\,,
\end{equation}
which means that the zitterbewegung is always perpendicular to the group
velocity of the wave packet. Let us therefore concentrate on the case
$k_{0x}=0$. By expanding the exponential containing the trigonometric functions
one derives 
\begin{eqnarray}
\langle\psi|x_{H}(t)|\psi\rangle  & = & \frac{1}{2k_{0y}}
\left(1-e^{-d^{2}k_{0y}^{2}}\right)\nonumber\\
 & -  & \frac{1}{2k_{0y}}\sum_{n=0}^{\infty}\Biggl[
\frac{\left(dk_{0y}\right)^{2(n+1)}}{n!(n+1)!}\nonumber\\
 & & \quad\quad\cdot\int_{0}^{\infty}dq q^{2n+1}e^{-q^{2}}
\cos\left(\frac{2\alpha q}{\hbar d}t\right)\Biggr]\,.
\end{eqnarray}
Thus, the amplitude of the zitterbewegung is proportional to the
wave length of the electron motion perpendicular to it, and the oscillatory
zitterbewegung changes its sign if the translational motion is reversed.
If the product $dk_{0y}$ is not too large, $dk_{0y}\lesssim 1$, only low
values of the summation index $n$ lead to substantial contributions, and the
Gaussian factor in the integrand suppresses contributions from large values
of $q$. Thus, a typical scale of this integration variable is leading to
sizable contributions is $q\approx 1/\sqrt{2}$. Thus, a typical time scale
in the integrand is $T=\sqrt{2}\pi\hbar d/\alpha$, and when averaging the
zitterbewegung over times scales significantly larger than $T$, the
cosine term drops giving
\begin{equation}
\overline{\langle\psi|x_{H}(t)|\psi\rangle}
=(1/2k_{0y})\left(1-\exp(-d^{2}k_{0y}^{2})\right)\,,
\end{equation} 
i.e. the time-averaged guiding center of the wave packet is shifted 
perpendicular to its direction of motion. Note that the zitterbewegung is
absent for $k_{0y}=0$ \cite{Huang52}.

In the opposite case $dk_{0y}\gg 1$ the Gaussian approaches a 
$\delta$-function. In this limit one finds (for $k_{0x}=0$)
\begin{equation}
\langle\psi|x_{H}(t)|\psi\rangle=
(1/2k_{0y})\left(1
-\cos\left(2\alpha k_{0y}t/\hbar\right)\right)\,.
\end{equation}
Here the frequency of the zitterbewegung is $\Omega=2\alpha k_{0y}/\hbar$, 
and the guiding center of the wave packet is also
shifted in the direction perpendicular to its group velocity. Note
that $\hbar\Omega$ is the excitation energy between the two branches of the
Rashba Hamiltonian ${\cal H}$ at a given momentum $\vec k=k_{0y}\vec e_{y}$.

\begin{figure}
\centerline{\includegraphics[width=8cm]{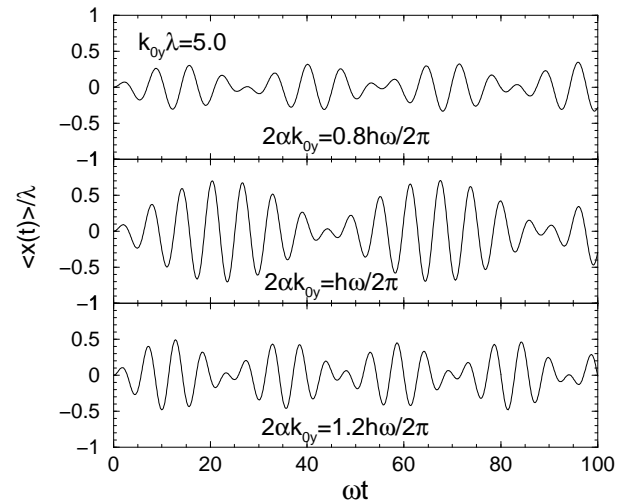}} 
\caption{Zitterbewegung of an electron in a harmonic quantum wire
perpendicular to the wire direction. The wave number $k_{0y}$ for the motion 
along the wire is $k_{0y}\lambda=5$. The amplitude of the zitterbewegung 
is maximal at the resonance $2\alpha k_{0y}=\hbar\omega$
(middle panel).
\label{fig1}}
\end{figure}

\begin{figure}
\centerline{\includegraphics[width=8cm]{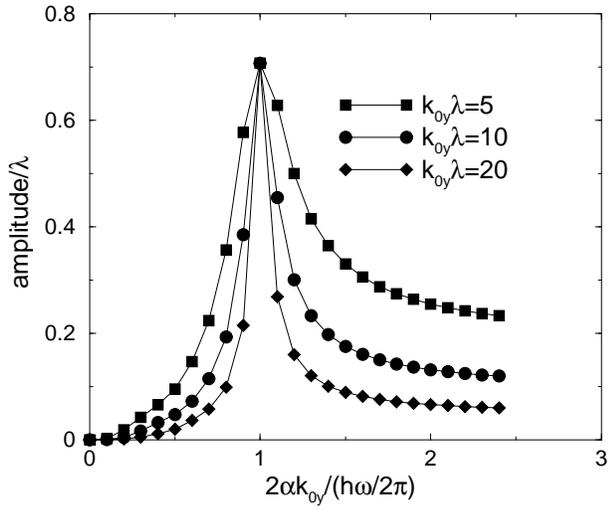}} 
\caption{Amplitude of the zitterbewegung perpendicular to the wire direction
as a function of $\Omega/\omega=2\alpha k_{0y}/\hbar\omega$ for different 
values of the wave number $k_{0y}$ along the wire.
\label{fig2}}
\end{figure}

The zitterbewegung of an electron in a quantum well as described above
is naturally accompanied by a broadening of the wave packet, where the
dominant contribution stems from the dispersive effective-mass term in the
Hamiltonian. Such a broadening might pose an obstacle for experimentally
detecting the zitterbewegung. However, the broadening can be efficiently 
suppressed and limited if the electron moves along a quantum wire.
In fact, the motion of electrons in quantum wells is generally under better 
control if additional lateral confinement is present. 
We therefore consider a harmonic
quantum wire along the $y$-direction described by
${\cal H}=\vec p^{2}/2m+m\omega^{2}x^{2}/2+{\cal H}_{R}$, where the frequency
$\omega$ parametrizes the confining potential perpendicular to the wire
\cite{wire,Governale02}.
For this case exact analytical progress as above does not seem to be possible,
and we therefore follow a numerical approach combined with an approximate
analytical study. To be specific, we
consider an electron with a given momentum $k_{0y}$ along the wire and
injected initially into the lowest subband of the confining potential
with the spin pointing upwards 
along the $z$-direction, i.e. the initial wave function for the $x$-direction
is a Gaussian whose width is determined by the characteristic length
$\lambda=\sqrt{\hbar/m\omega}$ of the harmonic confinement. In our numerical 
simulations of the exact time evolution we find again a zitterbewegung
perpendicular to the electron motion along the wire with the width
of the wave function across the wire being limited by the confining potential.
Moreover, the amplitude of the zitterbewegung becomes maximal if the
resonance condition $|\hbar\Omega|=|2\alpha k_{0y}|=\hbar\omega$ is fulfilled.
This general finding is illustrated in Fig.~\ref{fig1} where the wave number
along the wire is fixed to be $k_{0y}\lambda=5$ and the Rashba parameter
$\alpha$ is varied around the resonance condition. Equivalent observation are 
made if the Rashba coupling is fixed while the wave number $k_{0y}$ is varied.
In Fig.~\ref{fig2} we have plotted the amplitude of the zitterbewegung
as a function of $\Omega/\omega=2\alpha k_{0y}/\hbar\omega$ for different 
values of the wave number $k_{0y}$ along the wire. In this range of parameters,
the resonance bocomes narrower with increasing $k_{0y}$, while
its maximum value is rather independent of this quantity and 
remarkably well described by $\lambda/\sqrt{2}$. 

A qualitative explanation
for this resonance can be given by writing the Hamiltonian in the form
${\cal H}={\cal H}_{0}+{\cal H}_{1}$ with 
${\cal H}_{0}=\hbar\omega(a^{+}a+1/2)+\hbar^{2}k_{0y}^{2}/2m
+\alpha k_{0y}\sigma^{x}$,
${\cal H}_{1}=-i\sqrt{\hbar m\omega/2}(\alpha/\hbar)(a-a^{+})\sigma^{y}$,
and $a$, $a^{+}$ being the usual harmonic climbing operators
\cite{Governale02}. The 
zitterbewegung is induced by the perturbation ${\cal H}_{1}$ which can act most
efficiently if the unperturbed energy levels of ${\cal H}_{0}$ are 
degenerate having opposite spins. This is the case at 
$|2\alpha k_{0y}|=\hbar\omega$.

Another way to understand this resonance condition is to consider a
truncated model where the Hamiltonian has been projected onto the lowest
to orbital subbands, an approximation which is known to give very reasonable
results for the low-lying energy spectrum of the wire \cite{Governale02}. 
For a given wave number $k_{0y}$ the truncated
Hilbert space is spanned by the states $|0,\uparrow\rangle$, 
$|0,\downarrow\rangle$, $|1,\uparrow\rangle$, $|1,\downarrow\rangle$, 
where the arrows denote the spin state with respect to the $z$-direction,
and $0$ and $1$ stand for the ground state and the first excited state of
the harmonic potential, respectively. When applying the transformation
\begin{equation}
U=\frac{1}{\sqrt{2}}\left(
\begin{array}{cccc}
1 & 1 & 0 & 0  \\
0 & 0 & 1 & -1 \\
1 & -1 & 0 & 0 \\
0 & 0 & 1 & 1 
\end{array}
\right)
\end{equation}
the projected Hamiltonian and in turn the time evolution operator become
block-diagonal,
\begin{equation}
Ue^{-\frac{i}{\hbar}Ht}U^{+}=\left(
\begin{array}{cc}
h_{+}(t) & 0 \\
0 & h_{-}(t)
\end{array}
\right)
\end{equation}
where
\begin{eqnarray}
h_{\pm}(t) & = & 
\left({\bf 1}\cos\left(\frac{1}{\hbar}\mu_{\pm}t\right)
-i\frac{\vec\mu_{\pm}}{\mu_{\pm}}\cdot\vec\sigma
\sin\left(\frac{1}{\hbar}\mu_{\pm}t\right)\right)\nonumber\\
 & & \cdot\exp\left(\frac{i}{\hbar}\left(\hbar\omega
+\frac{\hbar^{2}k_{0y}^{2}}{2m}\right)t\right)
\end{eqnarray}
and $\vec\mu_{\pm}=(\pm\alpha\sqrt{m\omega/2\hbar},0,
-\hbar\omega/2\mp\alpha k_{0y})$.
Now using 
\begin{equation}
UxU^{+}=\frac{\lambda}{\sqrt{2}}\left(
\begin{array}{cc}
0 & \sigma^{x} \\
\sigma^{x} & 0 
\end{array}
\right)
\end{equation}
one obtains for the initial state $|0,\uparrow\rangle$ the
following time-dependent expectation value
\begin{eqnarray}
\langle 0,\uparrow|x_{H}(t)|0,\uparrow\rangle & = &
\frac{\lambda}{\sqrt{2}}\frac{\mu_{+}^{x}\mu_{-}^{z}+\mu_{-}^{x}\mu_{+}^{z}}
{\mu_{+}\mu_{-}}\nonumber\\
 & & \cdot\sin\left(\mu_{+}t/\hbar\right)
\sin\left(\mu_{-}t/\hbar\right)\,.
\end{eqnarray}
At resonace, $\hbar\omega=\pm2\alpha k_{0y}$, we have $\mu_{\mp}^{z}=0$,
and if $\mu_{\pm}^{x}$ can be neglected compared to $\mu_{\pm}^{z}$
(which is the case for large enough $k_{0y}$) the amplitude of the 
zitterbewegung is approximately $\lambda/\sqrt{2}$, in accordance with the
numerics.

We note that the above
resonance condition can be reached by tuning either the Rashba coupling
or the wave vector of the electron injected into the wire. For instance, for
a wire of width $\lambda=50{\rm nm}$ and the band mass of InAs being
$0.023$ in units of the bare electron mass, one has 
$\hbar\omega\approx 1{\rm meV}$. Typical values for the 
Rashba parameter $\alpha$ in InAs
are of order $10^{-11}{\rm eVm}$ 
\cite{Nitta97,Engels97,Heida98,Hu99,Grundler00,Sato01,Hu03}
leading to a wave length
$\lambda_{0y}=2\pi/k_{0y}\approx 100{\rm nm}$ at resonance. For GaAs, the
band mass is larger while the Rashba coefficient is typically an order of
magnitude smaller than in InAs \cite{Miller03},
giving a wave length of a few ten nm at resonance for $\lambda=50{\rm nm}$.

We are propose that electron zitterbewegung in semiconductor nanostructures
as described above can be experimentally observed using high-resolution
scanning-probe microscopy 
imaging techniques as developed and discussed in 
Refs.~\cite{Topinka00,LeRoy03}. As a possible setup, a tip sufficiently
smaller than the width of the wire can be moved along the wire and 
centered in its middle. For an appropriate biasing of the tip, the 
electron density at its location is depleted leading to a reduced conductance
of the wire. Since the amplitude of the
zitterbewegung reflects the electron density in the center of the wire, 
the zitterbewegung will induce beatings in the wire conductance as
a function of the tip position. These beatings are most pronounced at the
resonance, see Fig.~\ref{fig1}. Note that the oscillations shown there as
a function of time can be easily converted to the real-space $y$-coordinate
by multiplying the abscissa by $\hbar k_{0y}/m$. Generally we expect spin-orbit
effects in STM experiments to be more pronounced in the presence of
additional confinement such as in a quantum wire.

Moreover let us consider spin-orbit coupling of the Dresselhaus type
\cite{Dresselhaus55,Dyakonov86,Bastard92},
\begin{equation}
{\cal H}_{D}=(\beta/\hbar)\left(p_{y}\sigma^{y}-p_{x}\sigma^{x}\right),
\label{dressel}
\end{equation}
which is due to the bulk-inversion asymmetry in zinc-blende semiconductors,
and the coefficient $\beta$ is determined by the well width along with 
a material constant \cite{Dyakonov86,Bastard92}.
Here the components of the time-dependent position operator 
of an electron in a quantum well read
\begin{eqnarray}
x_{H}(t) & = & x(0)+ \frac{p_{x}}{m}t
-\frac{p_{y}}{p^{2}}\frac{\hbar}{2}\sigma^{z}
\left(1-\cos\left(\frac{2\beta p}{\hbar^{2}}t\right)\right)\nonumber\\
 & + & \frac{p_{x}}{p^{3}}\frac{\hbar}{2}
\left(p_{y}\sigma^{y}-p_{x}\sigma^{x}\right)
\left(\frac{2\beta p}{\hbar^{2}}t
-\sin\left(\frac{2\beta p}{\hbar^{2}}t\right)\right)\nonumber\\
 & - & \frac{1}{p}\frac{\hbar}{2}\sigma^{x}
\sin\left(\frac{2\beta p}{\hbar^{2}}t\right)\,,\\
y_{H}(t) & = & y(0)+ \frac{p_{y}}{m}t
+\frac{p_{x}}{p^{2}}\frac{\hbar}{2}\sigma^{z}
\left(1-\cos\left(\frac{2\beta p}{\hbar^{2}}t\right)\right)\nonumber\\
 & + & \frac{p_{y}}{p^{3}}\frac{\hbar}{2}
\left(p_{y}\sigma^{y}-p_{x}\sigma^{x}\right)
\left(\frac{2\beta p}{\hbar^{2}}t
-\sin\left(\frac{2\beta p}{\hbar^{2}}t\right)\right)\nonumber\\
 & + & \frac{1}{p}\frac{\hbar}{2}\sigma^{y}
\sin\left(\frac{2\beta p}{\hbar^{2}}t\right)\,.
\end{eqnarray}
In much the same way as above, these expressions lead to a zitterbewegung of
electronic wave packets in a direction perpendicular to their group velocity.
If both Rashba and the Dresselhaus spin-orbit
coupling are present, the directional dependence of the zitterbewegung
is more complicated, which can be understood  terms of the
anisotropic dispersion relations arising in this case \cite{Schliemann03a}.
The corresponding
expressions are rather lengthy and shall not be detailed here. 
However, in the case where the Rashba coefficient is tuned to be equal in
magnitude to the Dresselhaus term, $\alpha=\pm\beta$, the zitterbewegung is 
absent. This is due to the additional conserved quantity which arises
at this point and cancels the many effects of spin-orbit coupling
\cite{Schliemann03b,Schliemann03a}.

Finally we mention that similar expressions  can be derived for the case
of heavy holes in the p-type valence band of III-V 
semiconductors (as opposed to s-type electrons studied so far)
being subject to spin-orbit coupling
due to structure-inversion asymmetry \cite{Winkler00,Gerchikov92}.
Again, the zitterbewegung of a wave packet with its spin pointing initially 
in the $z$-direction is perpendicular to the group velocity.

In conclusion, we have studied zitterbewegung of electronic wave packets in 
III-V zinc-blende semiconductor quantum wells in the presence of
spin-orbit coupling of the Rashba and Dresselhaus type.
Our results suggest the possibility of 
a direct experimental proof of this oscillatory motion due to relativistic 
effects, confirming a long-standing theoretical prediction. Similar 
results can be derived for the case of heavy holes in
quantum wells under the influence of Rashba spin-orbit coupling.
A very promising route for such experiments are high-resolution
imaging techniques developed recently \cite{Topinka00,LeRoy03}.
If the spin of the electron
is initially aligned along the $z$-direction, the zitterbewegung is always
perpendicular to the group velocity of the wave packet.
For possible experiments quantum wires are particularly attractive.  
For this case we find a resonance condition maximizing the 
zitterbewegung. This resonance can be reached by either tuning the
Rashba coupling or the electron velocity along the wire.

This work was supported by the NCCR Nanoscience, the Swiss NSF, DARPA, ARO,
ONR, and the EU Spintronics RTN.


\end{document}